\newif\ifEPRINT \EPRINTtrue
\ifx\undefined\ifEPRINT
\documentstyle[aps,preprint,epsf]{revtex}
\else
\documentstyle[aps,multicol,epsf]{revtex}
\fi
\newcommand{\bold}[1]{\mbox{\boldmath $#1$}}    

\newcommand{\sLambda}{{\scriptscriptstyle \Lambda}}
\def\nuc#1#2{\relax\ifmmode{}^{#1}{\protect\text{#2}}\else${}^{#1}$#2\fi}
\def\C12{\nuc{12}{C}}
\newcommand{\VEV}[1]{<{#1}>} 
\newcommand{\lam}{$\Lambda(1405)$}
\newcommand{\beqar}{\begin{eqnarray}}
\newcommand{\eeqar}{\end{eqnarray}}
\newcommand{\beq}{\begin{equation}}
\newcommand{\eeq}{\end{equation}}
\newcommand{\raf}[1]{(\ref{#1})}

\newcommand{\geteps}[2]{\epsfxsize=#1in\epsfbox{#2}}    
\newcommand{\getps}[1]{%
        \begin{center}\hspace*{-0.0cm}\geteps{3}{PS/#1}\end{center}}
\newcommand\Figampl{%
\begin{figure}
\getps{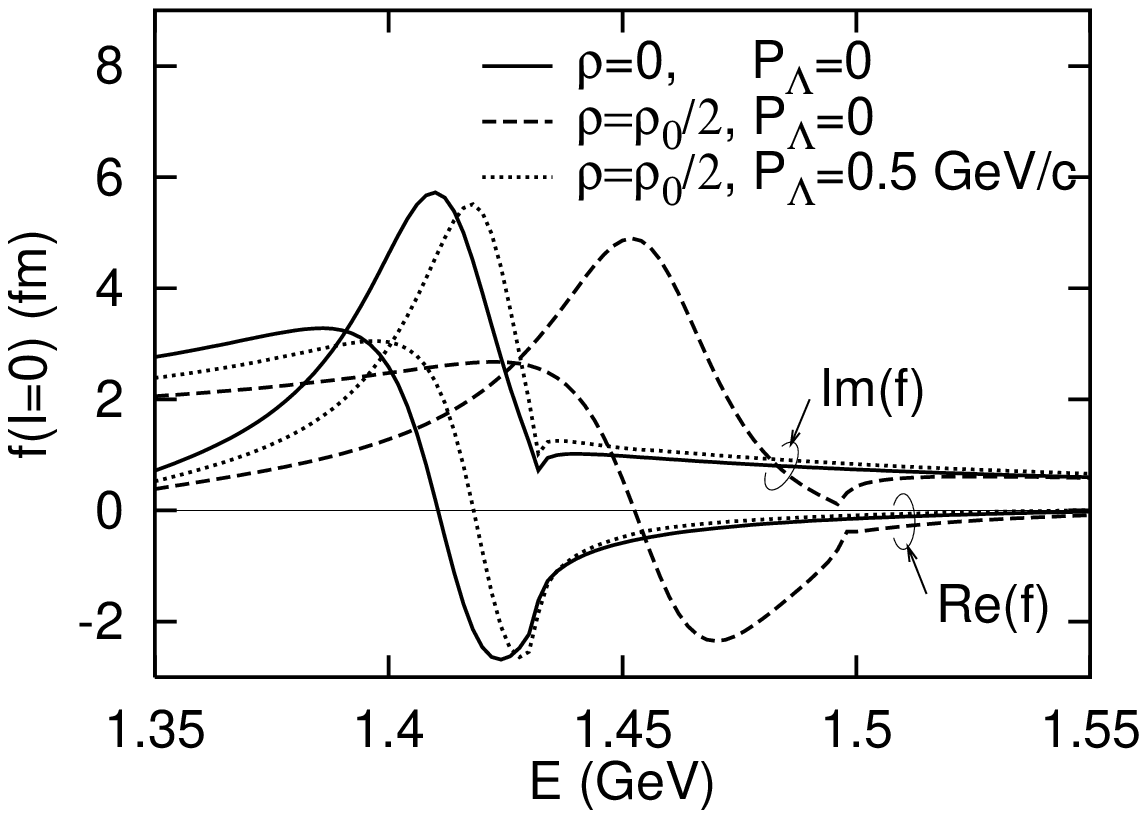}
\caption{Scattering Amplitude of $\bar{K}N$ in Nuclear Medium.
Energy dependence of the scattering amplitude in $I=0, \bar{K}N$ channel.
Solid, dashed, and dotted lines show the calculated results
at $(\rho,P_\sLambda)=(0,0), (\rho_0/2,0),$ and $(\rho_0/2,0.5 {\rm GeV/c})$,
respectively.}
\label{fig:ampl}
\end{figure}}
\newcommand\Figbrd{%
\begin{figure}
\getps{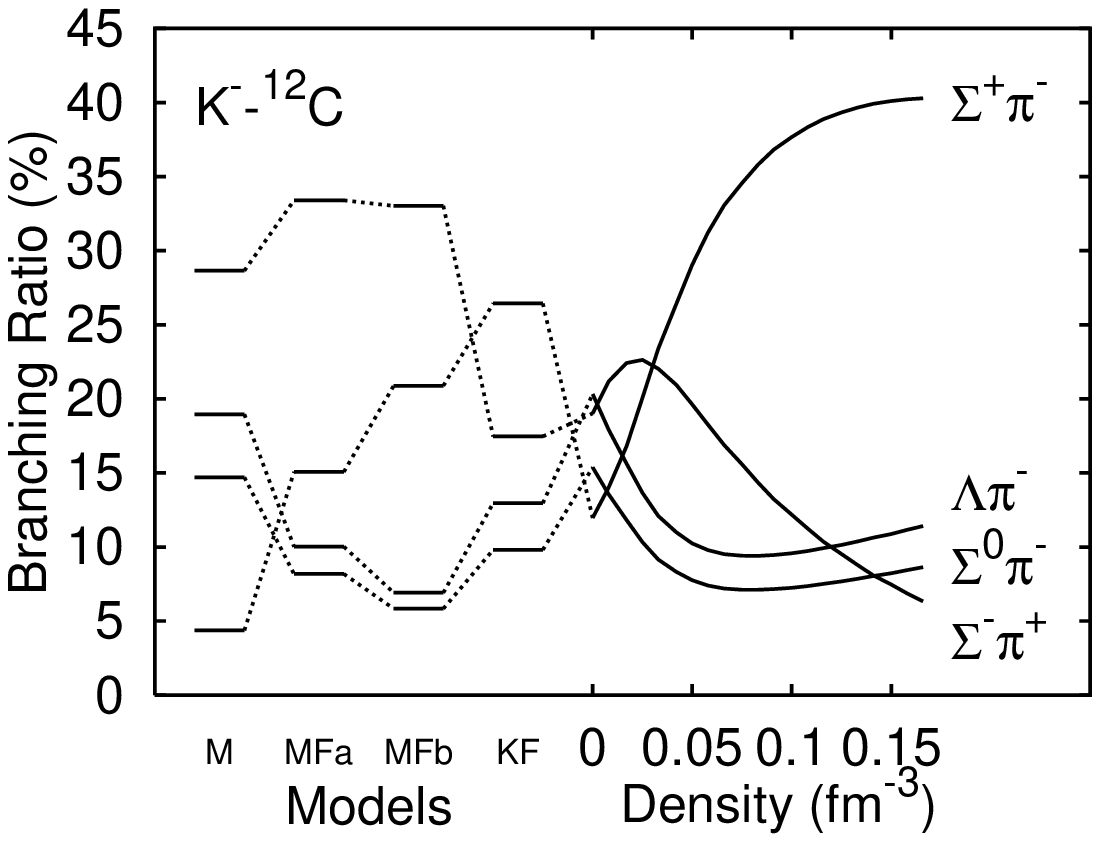}
\noindent
\caption{Model dependence of the branching ratio.
Branching ratios with models MFa, MFb are calculated through
the Fermi smeared $T$-matrix with binding energy correction,
which is density independent.
In the case of model KF,
we show the density dependence of the branching ratios
(right solid curves) and their average based on the probability distribution
shown in Fig.~\protect\ref{fig:distc}.}
\label{fig:brd}
\end{figure}}
\newcommand\Figpotc{%
\begin{figure}
\getps{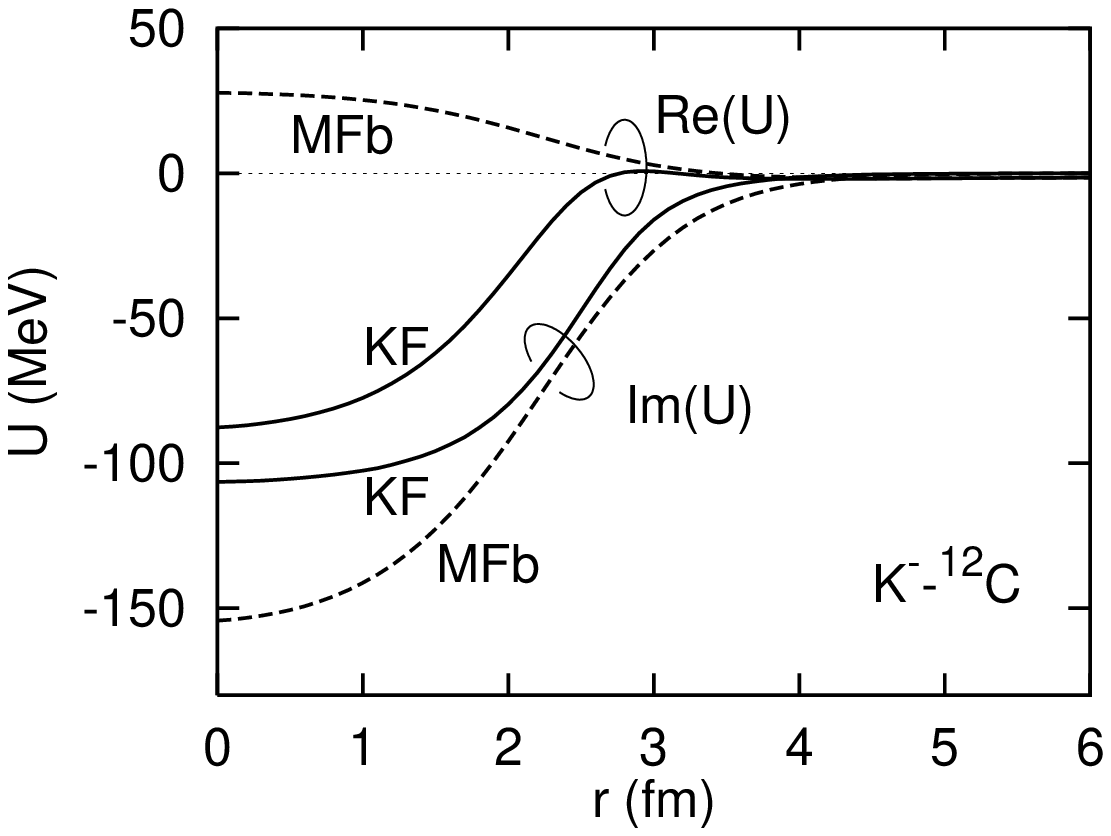}
\caption{Optical potential $U$ between $K^-$ and \C12.
Dashed and solid lines represent the potential calculated with
models of MFb and KF, respectively.
We have used the lowest order impulse approximation Eq.~\protect\raf{eq:opt}. }
\label{fig:potc}
\end{figure}}
\newcommand\Figdistc{%
\begin{figure}
\getps{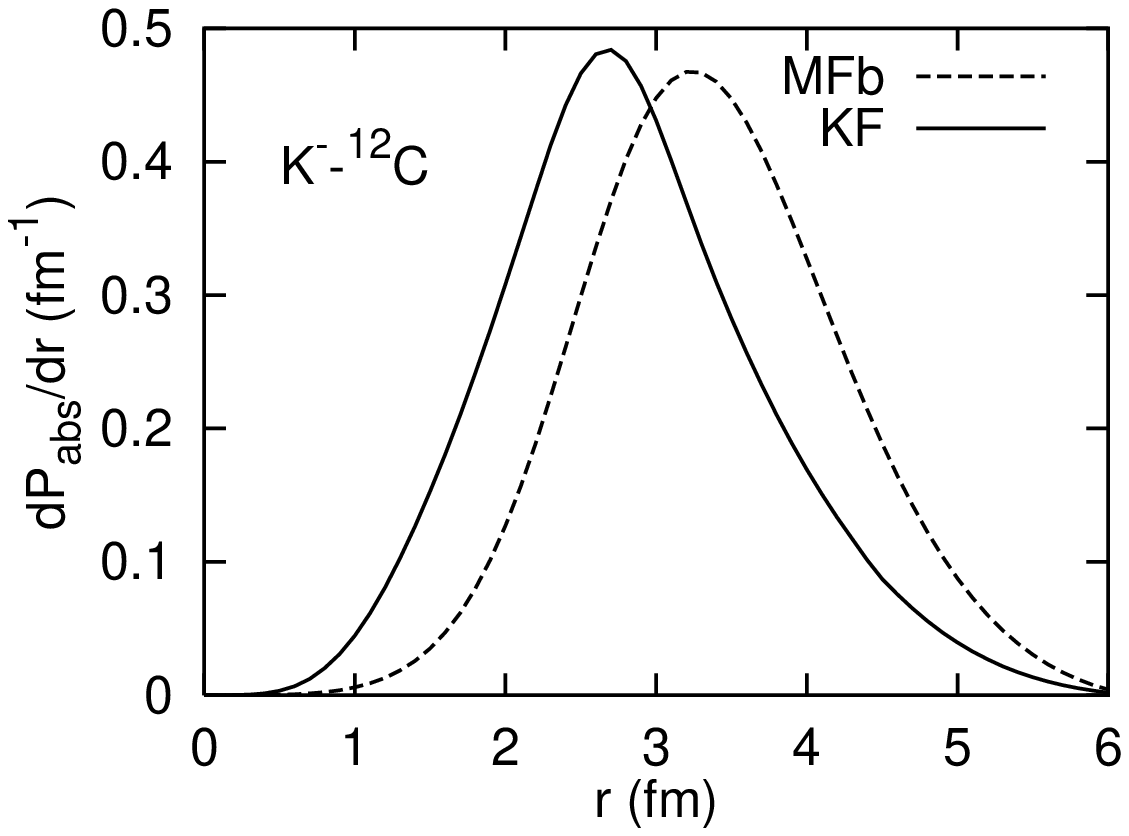}
\noindent
\caption{Absorption point distribution of $K^-$ on \C12.
Dashed and solid lines represent the calculated results with models
of MFa and KF, respectively.
The absorption probability is assumed to be proportional to
$-Im(U)\, |\psi(r)|^2$, where $\psi(r)$ is 
the solution of the Schr\"odinger equation with the optical potential $U$. }
\label{fig:distc}
\end{figure}}
\newcommand\Figstkcm{%
\begin{figure}
\getps{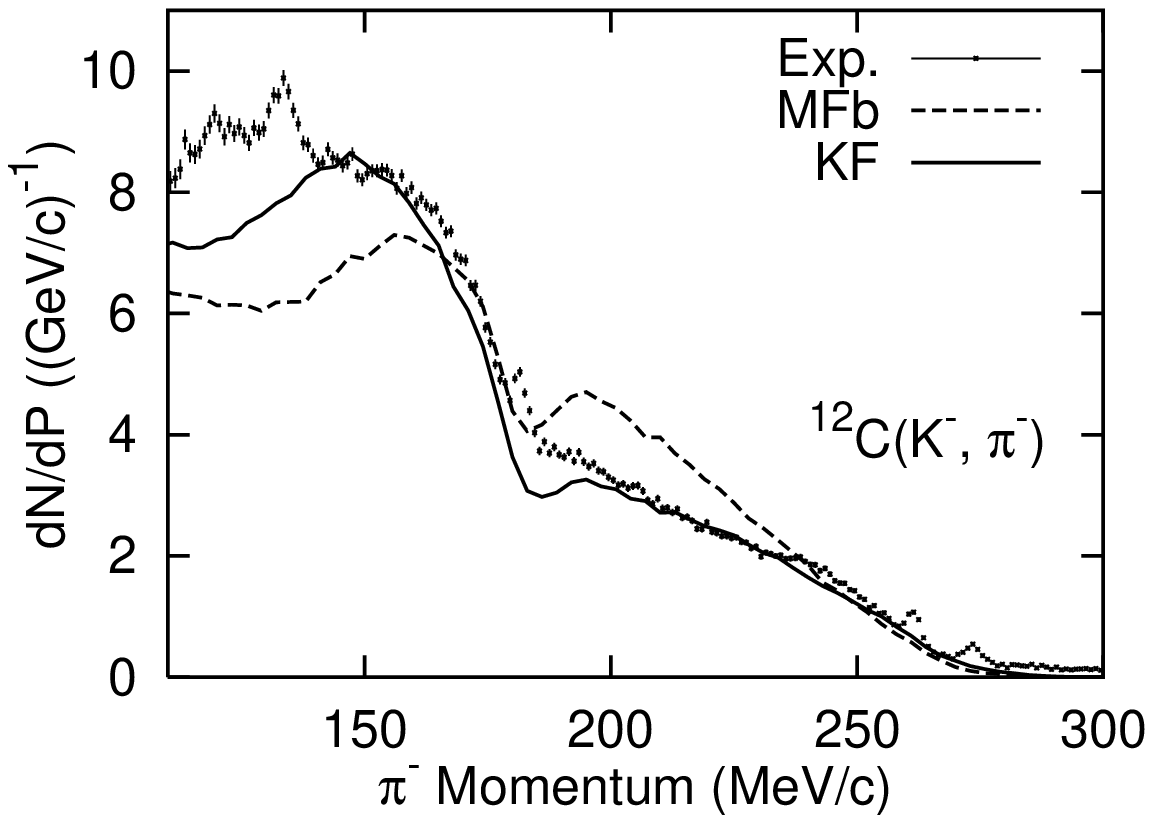}
\caption{$\pi^-$ momentum spectrum from $K^-$ absorption at rest on \C12.
Dashed and solid lines show the calculated results with models MFb and KF,
respectively.
Experimental data (point with error bar) is taken from \protect\cite{Tamura89}.
Since the absolute values are not shown in \protect\cite{Tamura89}, 
we have assumed that
$dN/dp|_{\rm exp.}(({\rm GeV/c})^{-1})={\rm Count/bin}|_{\rm exp.}/800$. }
\label{fig:stkcm}
\end{figure}}
\newcommand\Figstkcmcmp{%
\begin{figure}
\getps{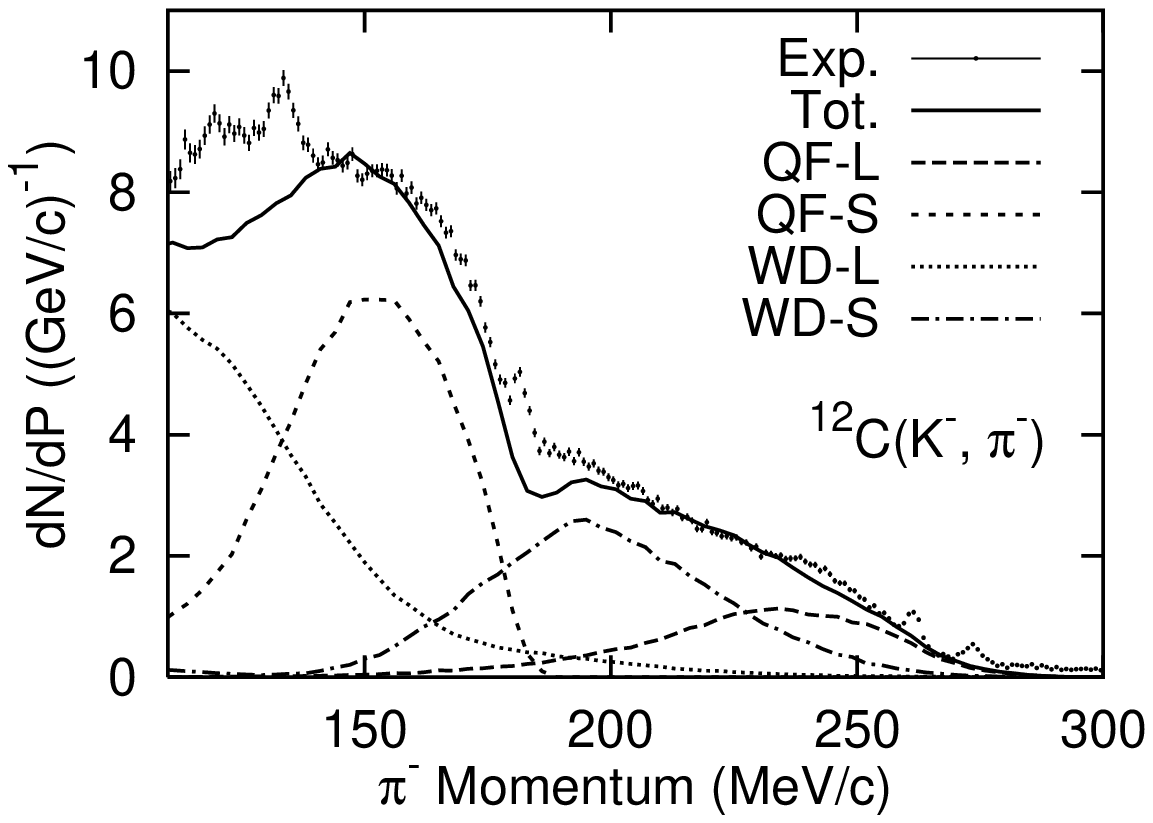}
\caption{Components in $\pi^-$ momentum spectrum
        from $K^-$ absorption at rest on \C12. 
Component dissolution of $\pi^-$ spectrum shown in Fig.~\protect\ref{fig:stkcm}
in the model KF.
Experimental data (point with error bar) is taken from \protect\cite{Tamura89}.
Solid line represents the total yield.
Long and short dashed lines show quasi-free $\Lambda$ and $\Sigma$ productions,
respectively.
Dotted and dot-dashed lines show the contributions of
$\Lambda$ and $\Sigma$ weak decays, respectively.
It is noteworthy that the $\Lambda$ weak decay component includes 
$\Sigma^0$ sequential decay and other $\Sigma$ conversion,
thus it has a longer tail in high momentum region 
compared with the estimate of Ref.\protect\cite{Tamura89}.}
\label{fig:stkcmcmp}
\end{figure}}
\newcommand\Figexpfit{%
\begin{figure}
\getps{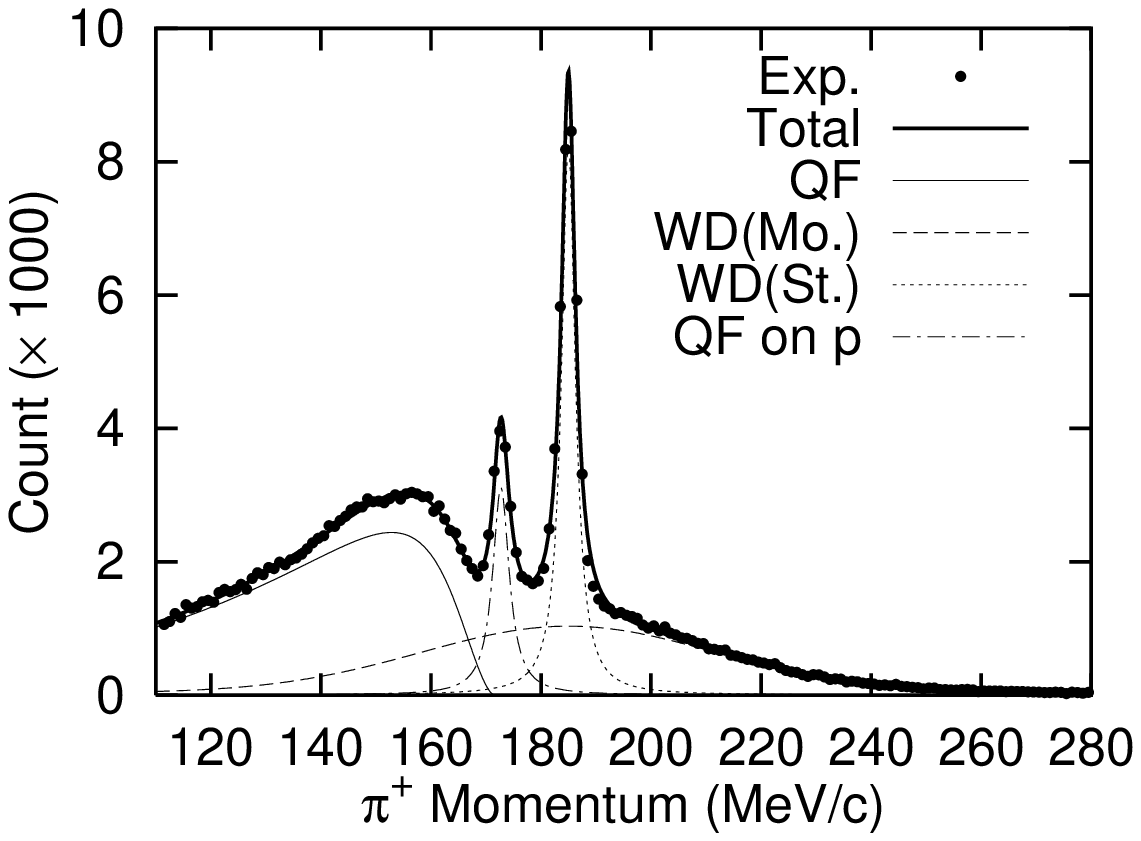}
\caption{%
The fitted spectrum to \C12$(K^-_{\rm stopped},\, \pi^+)$ data
in Ref.~\protect\cite{Kubota96}.
The thick sold curve show the fitted total spectrum to the experimental
data shown by dots.
The sharp peak at around 185 MeV/c comes form the weak decay of $\Sigma^+$ 
which stopped in the scintillator (fitted by the dotted curve),
and that at around 172 MeV/c corresponds to the free $\pi^+$
production on the protons in the plastic scintillator, 
$K^-p \to \Sigma^-\pi^+$ (dash-dotted curve).
The thin sold and dashed curves show the contribution from 
quasi-free $\pi^+$ production on \C12
and weak decay of moving $\Sigma^+$, respectively.
}
\label{fig:expfit}
\end{figure}}
\newcommand\Figstkcp{%
\begin{figure}
\getps{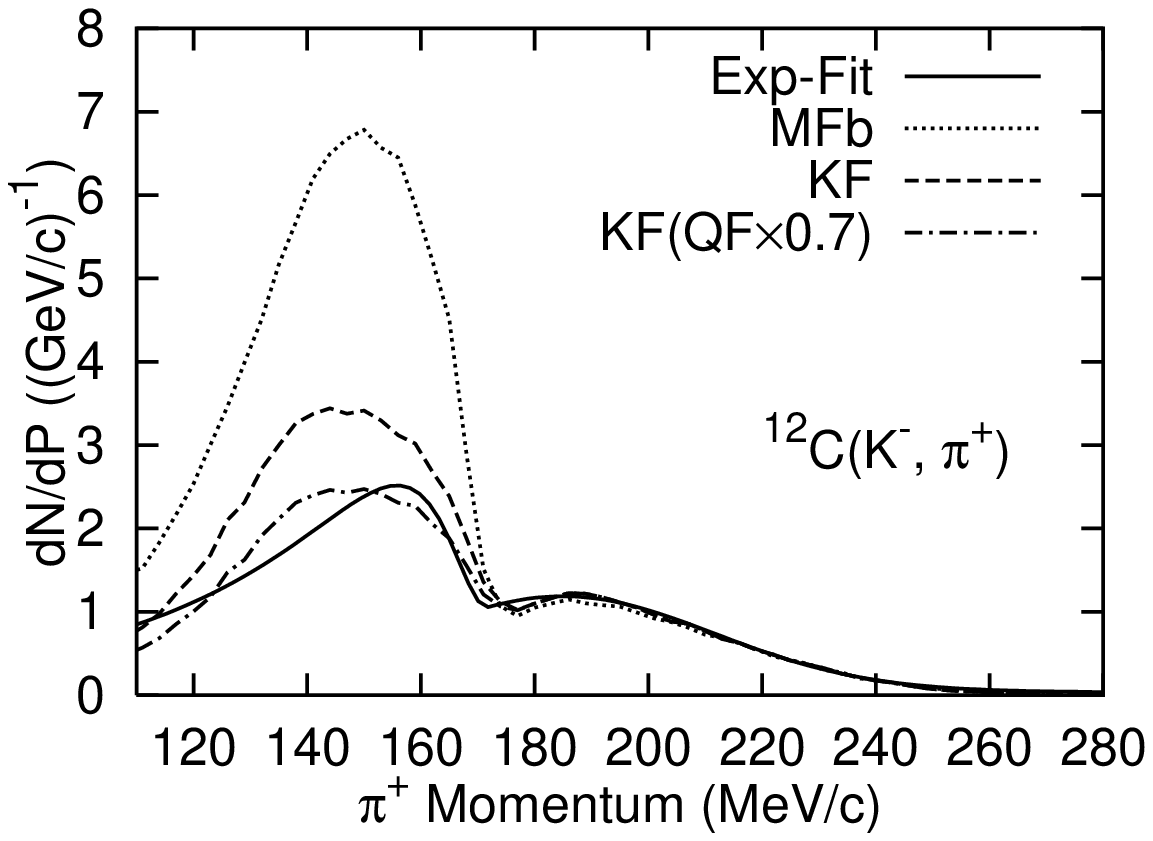}
\caption{%
$\pi^+$ momentum spectrum from $K^-$ absorption at rest on \C12.
Dotted and dashed lines show the calculated results with models MFb and KF,
respectively.
Our fitting result to the experimental data~\protect\cite{Kubota96}
is shown by the solid curve.
We have chosen the normalization factor
$dN/dp|_{\rm exp.}(({\rm GeV/c})^{-1})={\rm Count/bin}|_{\rm exp.}/1300$,
by requiring that the calculated weak decay tails agree with the fitting
result to the experimental data. 
This normalization factor corresponds to the number of experimentally
detected events of $1.3 \times 10^6$.
The dash-dotted line shows the modified calculated results of model KF,
with a reduction factor 0.7 for quasi-free $\pi^+$.
}
\label{fig:stkcp}
\end{figure}}
\newcommand\TableA{%
\begin{table}[hbt]
\caption{Average branching ratio in $K^-$ absorption at rest on $^{12}$C.
Branching ratios in models of M, MFa, MFb, KF are shown
in comparison with the experimentally estimated one~\protect\cite{Wilquet77}.}
\label{table:branch}
\begin{tabular}{lrrrrr}
 & M & MFa & MFb & KF & Exp. \\
\hline\hline
$K^-p \to \Lambda\pi^0$ &  9.5 &  5.0 &  3.5 &  6.5 &  4.4 \\
$K^-n \to \Lambda\pi^-$ & 18.9 & 10.0 &  6.9 & 13.0 &  8.7 \\
$K^-p \to \Sigma^-\pi^+$& 28.7 & 33.4 & 33.0 & 17.5 & 16.8 \\
$K^-p \to \Sigma^0\pi^0$&  9.2 & 20.1 & 24.0 & 17.0 & 25.7 \\ 
$K^-p \to \Sigma^+\pi^-$&  4.4 & 15.1 & 20.9 & 26.4 & 37.7 \\ 
$K^-n \to \Sigma^-\pi^0$& 14.7 &  8.2 &  5.8 &  9.8 &  3.3 \\
$K^-n \to \Sigma^0\pi^-$& 14.7 &  8.2 &  5.8 &  9.8 &  3.3 \\
\end{tabular}
\end{table}}
\newcommand\TableB{%
\begin{table}[hbt]
\caption{
Extracted multiplicity of escaped quasi-free $\pi^+$ and $\Sigma^+$
particles.
Multiplicities in model M, MFb, and KF are compared with those
estimated from the experimental data.
In the models MFb and KF, effects of $\pi$ rescattering and absorption
as well as $\Sigma$ conversion are included
within the Monte-Carlo simulation used in this work.
We have assumed that the number of events is $1.3 \times 10^6$
in the experiment.
}
\label{table:multi}
\begin{tabular}{lccccc}
                & M     & MFb   & KF    & Exp.  & (Counts)              \\
\hline\hline
QF $\pi^+$ (A)  & 0.287 & 0.298 & 0.153 & 0.103 & ($1.33 \times 10^5$)  \\
$\Sigma^+$ (B)  & 0.044 & 0.137 & 0.143 & 0.184 & ($2.40 \times 10^5$)  \\
Ratio (A/B)     & 6.558 & 2.180 & 1.075 & 0.554 &                       \\
\end{tabular}
\end{table}}

        \ifx\undefined\ifEPRINT
\newcommand{\NARROWTEXT}{}
\newcommand{\WIDETEXT}{}
\newcommand{\TABLEA}{%
        \begin{center}\fbox{Table ~\ref{table:branch}}\end{center}}
\newcommand{\TABLEB}{%
        \begin{center}\fbox{Table ~\ref{table:multi}}\end{center}}
\newcommand{\figampl}{\figposition{ampl}}
\newcommand{\figpotc}{\figposition{potc}}
\newcommand{\figdistc}{\figposition{distc}}
\newcommand{\figbrd}{\figposition{brd}}
\newcommand{\figstkcm}{\figposition{stkcm}}
\newcommand{\figstkcmcmp}{\figposition{stkcmcmp}}
\newcommand{\figexpfit}{\figposition{expfit}}
\newcommand{\figstkcp}{\figposition{stkcp}}
        \else
\newcommand{\NARROWTEXT}{\begin{multicols}{2} \global\columnwidth20.5pc}
\newcommand{\WIDETEXT}{\end{multicols} \global\columnwidth42.5pc}
\newcommand{\TABLEA}{%
	\noindent
	\begin{minipage}{8.5cm}\TableA\end{minipage}}
\newcommand{\TABLEB}{%
	\noindent
	\begin{minipage}{8.5cm}\TableB\end{minipage}}
\newcommand{\figampl}{\Figampl}
\newcommand{\figpotc}{\Figpotc}
\newcommand{\figdistc}{\Figdistc}
\newcommand{\figbrd}{\Figbrd}
\newcommand{\figstkcm}{\Figstkcm}
\newcommand{\figstkcmcmp}{\Figstkcmcmp}
\newcommand{\figexpfit}{\Figexpfit}
\newcommand{\figstkcp}{\Figstkcp}
        \fi

\begin{document}
\ifx\undefined\ifBIBTEX\else
\bibliographystyle{prsty}
\fi
\ifx\undefined\ifEPRINT \begin{titlepage} \fi
\preprint{HUPS-97-1, LBNL-40007}
\draft

\title{Branching ratio change in $K^-$ absorption at rest\\
	and the nature of the \lam\footnote{preprint HUPS-97-1, LBNL-40007}}

\author{Akira Ohnishi$^{1,2}$\thanks{e-mail: ohnishi@nucl.phys.hokudai.ac.jp},
		Yasushi Nara$^{1,3}$\thanks{e-mail: ynara@hadron03.tokai.jaeri.go.jp}, 
		and Volker Koch$^2$\thanks{e-mail: vkoch@nsdssd.lbl.gov}
	}

\address{1.\ Division of Physics, Graduate School of Science,
                Hokkaido University, Sapporo 060, Japan}

\address{2.\ Nuclear Science Division,
                Lawrence Berkeley National Laboratory,\\
                University of California, Berkeley, California 94720}
\address{3.\ Advanced Science Research Center,
                Japan Atomic Energy Research Institute,\\
                Tokai, Ibaraki 319-11, Japan}
\date{\today}

\maketitle

\begin{abstract}
We investigate in-medium corrections to the branching ratio in 
$K^-$ absorption at rest and their effect on the charged pion $\pi^\pm$ 
spectrum. The in-medium corrections are due to Pauli blocking, which arises 
if the \lam\ is assumed to be a $\bar{K}$-nucleon bound state and leads to 
a density and momentum dependent mass shift of the \lam. 
Requiring that the optical potential as well as the branching ratio are derived
from the same elementary $T$-matrix, we find that the in-medium corrected,
density dependent $T$-matrix gives a better description
of the $K^-$ absorption reaction
than the free, density-independent one.
This result suggests that the dominant component of the \lam\ wave function 
is the $\bar{K}N$ bound state. 
\end{abstract}


%
%
%

\ifx\undefined\ifEPRINT\end{titlepage}\fi

\NARROWTEXT


\section{Introduction}

%
%
The problems of the low-energy $K^-N$ interaction are longstanding.
It is well known that the s-wave $K^-$ nucleon scattering length
is repulsive at threshold\cite{Martin81} ($a_{K^-N} = -.15 \, $fm),
and a recent experiment clearly showed that the energy shift of
the 1s atomic orbit of kaonic hydrogen is positive
(i.e. $K^-p$ interaction is repulsive)~\cite{Ito-PhD,Iwasaki97}.
However, from a theoretical point of view it is more natural that the basic
$K^-N$ interaction is attractive. For instance, the leading order
(Weinberg-Tomazawa-type) term in the chiral expansion is attractive for the
$K^-N$ channel.

%
%
If the s-wave, isospin $I=0$ \lam\ resonance is a bound state
of $\bar{K}N$~\cite{DWR67,SW88},
it is possible that the actual $K^-p$ interaction is attractive 
although it appears repulsive in the scattering length 
and the energy shift of the kaonic hydrogen. 
This \lam\ resonance is clearly seen in $K^- p \rightarrow \Sigma \pi $
reactions, and couples strongly to the $K^- p $ channel. 
Since this \lam\ lies just below the $K^- p$ threshold,
scattering through this resonance gives rise to a repulsive contribution
to the scattering amplitude at threshold.
This effect is readily understood by analogy to the proton-neutron ($pn$) 
scattering; the interaction between $pn$ is attractive,
but the scattering length in the deuteron channel ($I=0, S=1$) is repulsive,
due to the existence of the deuteron as a bound state.
In nuclear matter, however,
the deuteron disappears, largely due to Pauli blocking,
and the true attractive nature of the $pn$ interaction emerges.
Similarly, the nature of the \lam\ 
--- a continuum bound state of $\bar{K}N$ or a genuine 3-quark state
--- can be studied by investigating the 
properties of the \lam\ in nuclear matter
since only the bound state component will be subject to Pauli-corrections.

%
%
One of the in-medium corrections suggested in the bound state picture
is the mass shift of \lam\,
and the resulting density dependence of the real part of
the $K^-$ optical potential in nuclear matter,
as extracted from the analysis of kaonic atoms \cite{FGB93}.
There one finds that the optical potential changes sign
--- from repulsive to attractive --- at rather low densities.
In the bound state picture of \lam, 
this density dependence can be simply understood
as the effect of the Pauli blocking of the proton inside the \lam,
which leads to an upwards shift of the \lam\ and thus
to an attractive potential~\cite{KochLam} (see also \cite{Weise}),
analogously to the deuteron and $pn$ interaction in medium as described before.
The resulting attractive $K^-$ potential has interesting
phenomenological consequences such as a possible condensate in neutron 
stars\cite{BLR94}.

%
%
While the $K^-$ optical potential is connected
with the {\em diagonal} matrix element of the elementary $T$-matrix
in the lowest order impulse approximation, 
the branching ratio from $K^-N$ to $\pi\Lambda$
and $\pi\Sigma$ is given by the {\em off-diagonal} matrix element
of the same $T$-matrix.
Since the $I=0$ amplitude (\lam\ channel) interferes
with that in the $I=1$ channel~\cite{DT60},
it is expected that the branching ratio changes considerably
in the nuclear medium if the mass of the \lam\ is shifted.
This in-medium branching ratio correction 
should be seen in the particle (pair) multiplicities
and in the pion spectra
from $K^-$ absorption reactions
at rest on nuclear targets, ${}^AZ(K^-_{\rm Stopped},\pi)$%
~\cite{Wilquet77,Outa94,Tamura89,Tamura88,Tamura94,Kubota96}, 
which contain various contributions of elementary processes, 
such as quasi free processes $K^-N \to Y \pi$,
and the weak decay of hyperons $Y \to \pi N$.
%
%
One of the pioneering work on the in-medium branching ratio
can be seen in the analyses of the bubble chamber data~\cite{Wilquet77}.
An in-medium branching ratio
for the primary interactions in \C12\ has been extracted   
by counting several kinds of charged particle pairs
observed in the bubble chamber.
Although some assumptions in went into this estimate, 
such as the conversion probability of $\Sigma$,
the extracted branching ratios are clearly different from the free one.
For example, the extracted in-medium branching ratio of the elementary process
$K^-p \to \Sigma^+\pi^-$ is much larger than that of $K^-p \to \Sigma^-\pi^+$,
while the latter channel dominates in free space. 
%
%
In recent counter experiments,
Outa et al. decomposed the experimental pion spectrum
into the contributions from various elementary channels in 
\nuc{4}{He}$(K^-_{\rm stopped},\pi^\pm)$ reactions
by using the delayed time gate technique~\cite{Outa94}.
In the case of heavier targets such as \C12, however, 
it is very difficult to decompose $\pi^\pm$ spectra
in an unambiguous way
solely based on experimental data.
Therefore, Tamura et al.~\cite{Tamura94} obtained
the spectral shapes of each contribution by  
carrying out a Monte Carlo simulation,
which took account of Fermi-motion of nucleons and energy loss of hyperons
in the target. The relative intensity of each component was then 
determined by fitting 
the sum of all contributions to the experimental data.

%
%
All these experimental data suggest the necessity of
the in-medium branching ratio corrections, 
although 
the final state interactions 
make it difficult
to extract the primary branching ratio.
Thus, it is required to
explain the experimental pion spectra themselves 
by means of theoretical models
which take the in-medium corrections and multi-step interactions into account.
In order to evaluate the final state multi-step interactions of
the primary hyperon and pion, 
cascade type models~\cite{Yariv79,Cugnon81,BUU,Giessen,NOHE97}
are considered to be the most reliable at present.
In these models, each elementary two-body collision process is treated
explicitly, which allows to include various multi-step processes.

%
%
The purpose of this paper is to elucidate the nature of \lam\ resonance
by analyzing the pion spectrum in $K^-$ absorption reactions at rest on nuclei.
We compare the calculated results based on two different scenarios
of the \lam\ resonance.
The first scenario assume that the \lam\ resonance is a genuine 3-quark
state, and, hence, is not subject to Pauli blocking by nucleons in nuclei.
In the second scenario,
we assume that \lam\ is a bound state of $\bar{K}N$,
and the Pauli blocking of nucleons shifts the mass of \lam\ upwards
in nuclear medium.
These two scenarios give different branching ratios
and different absorption point distributions, as well.
The first difference comes from the {\em off-diagonal} matrix elements.
The difference in the absorption point distribution on 
he other hand is due to a
different $K^-$ optical potential obtained in the bound state picture
\cite{KochLam}. It is, thus, related to the {\em diagonal} 
elements of the $T$-matrix.
The combined effects lead to significantly different pion spectra,
especially in the $(K^-,\pi^+)$ channel.
After taking into account final state interactions of
the produced hyperon and pion within the cascade model, 
it becomes possible to compare the calculated spectra
with experimental data.
From this comparison, 
we find that the interpretation of \lam\ as a 
$\bar{K}N$ bound state gives a better description of $K^-$ absorption
reactions at rest.

%
%

%
%
The necessity of the branching ratio modification in ${}^AZ(K^-,\pi)$
reactions has already been pointed out 
in Refs.~\cite{DMG,Harada,Rosental80,Bardeen71}.
These proposed  modifications come from the Fermi smearing and binding effects
and thus essentially smear and shift
both of the amplitudes in $I=0$ and $I=1$ channels
in a similar way. 
On the other hand, since  a mass shift of the \lam\ only affects the
$I=0$ channel, we expect significant differences between the branching 
ratio calculated with only Fermi-smeared amplitudes and those where in
addition the mass shift of the \lam\ is taken into account. 
Actually, we will show that the branching ratio strongly depends 
on the scenario of \lam\ in the case of $K^-$ particle absorption 
at rest on \C12.

%
%
The scenarios considered here, pure bound state on the one hand or pure
three-quark state on the other, are of course extreme. Most likely, the \lam\
will be a superposition of both. However, our results suggest that
the dominant component of the wave function of the \lam\
indeed is the $K^-p$ bound state. 
Typically in models which take such an admixture into account, the {\em bare}
\lam\ state has a mass much higher then $1405$~MeV
and it is the coupling to the $K^-p$ state which brings the mass down 
to the correct value~\cite{Masutani88,HAM95}.
In medium, these $K^-p$ states are also Pauli blocked leading again
to an upwards mass  shift of the \lam. Taking the mass of the bare \lam\ to
infinity in this approach would correspond the extreme bound state picture 
presented here.

Throughout this paper,
we assume that the in-medium corrections to the $I=1$
amplitude are negligible.

\section{The model of \lam}

In this section, we briefly summarize the description of \lam\ as
a bound state of $\bar{K}N$, and its modification in nuclear medium.
We follow the treatment of Ref.~\cite{KochLam}.

We consider two $I=0$ channels, $\bar{K}N$ and $\pi\Sigma$, 
to describe the \lam\ resonance.
Restricting ourselves to a non-relativistic treatment of the problem,
the mass and width of the \lam\ is obtained from the solution of the coupled
channel  Schr\"odinger equation,
\beq
\label{eq:Sch}
\nabla^2 \psi_i(r) + k_i^2 \psi_i(r)
         - 2 \mu_i \int V_{i,j}(r,r') \, \psi_j(r') d^3r'
                = 0\ ,
\eeq
where
$\psi_i(r)$ represents the relative wave function,
$\mu_i$ is the reduced mass,
and $k_i$ is the momentum in the center of mass system
for the channel $i=1(\bar{K}N), 2(\pi\Sigma)$.
Using a separable potential, the expression of the $T$-matrix is 
\beqar
\nonumber
V_{i,j}(k,k') &=&  g^2 C_{i,j} \,\, v_i(k) \, v_j(k')\\
\label{eq:seppot}
              &=& \frac{g^2}{\Lambda^2} C_{i,j} \Theta(\Lambda^2 - k^2 )
                  \Theta(\Lambda^2 - k'^2 )\ ,  \\
\label{eq:Tmat}
T_{i,j}(k,k',E)
        &=& g^2 \,v_i(k) \, v_j(k')
        \left[ (1- C \cdot G(E))^{-1} \cdot C \right]_{i,j}\ .
\eeqar
The propagator matrix $G$ is given by
\beqar
\label{eq:Prop}
G_{i,j} &=& diag(g_i) = \delta_{ij}\, g_i \ ,\\
\nonumber
g_i(E) &=&  g^2 \,
          \int \frac{d^3 p}{(2 \pi)^3} 
                \frac{v_i^2(p)}{E - m_i - M_i  - p^2/(2 \mu_i)} \\
\label{eq:prop}
        &=&\frac{1}{2 \pi^2} \frac{g^2}{\Lambda^2}
                \int_0^\Lambda
                        \frac{p^2 \, dp}{E - m_i - M_i - p^2/(2 \mu_i)}\ .
\eeqar
The scattering amplitude is directly related to the $T$-matrix by \cite{EK80}
\beq
\label{eq:ampl}
f_{i,j}(k,k') = - \frac{\sqrt{\mu_i\mu_j}}{2 \pi} T_{i,j}(k,k')\ .
\eeq
Finally, for the coupling matrix $C_{ij}$, we use the standard result
derived from the $SU(3)$ flavor symmetry (see e.g. Refs.\cite{DWR67,SW88})
\beq
\label{eq:couple}
C_{i,j} =
\begin{array}{cc}
\begin{array}{cc}
\bar{K}N & \pi \Sigma
\end{array} & \\
\left(
\begin{array}{cc}
-\frac{3}{2} & -\frac{\sqrt{6}}{4}\\
\\
-\frac{\sqrt{6}}{4} & -2
\end{array}
\right)
&
\begin{array}{c}
\bar{K}N\\
\\
\pi \Sigma
\end{array}
\end{array}
\quad .
\eeq
By choosing the parameters $\Lambda$ and $g$ 
appropriately, 
the scattering amplitude extracted from experimental
data by Martin~\cite{Martin81} can be reproduced
(see \cite{KochLam} for details).
In this paper,
we have used the parameters $\Lambda=0.78$ GeV and $g^2=17.9$, 
which are taken from~\cite{KochLam}.

In the momentum space representation,
the effect of the Pauli blocking can be readily included.
In the nuclear medium, the intermediate proton states with momenta
$p \leq k_f$ are forbidden, where $k_f$ is the Fermi momentum, which changes
the propagator in the $\bar{K}N$ channel to,
\beq
\label{eq:propmat}
g_1(E,k_f) = \frac{g^2}{\Lambda^2}
        \int_0^\Lambda \frac{d^3 p}{(2\pi)^3} \,
        \frac{\Theta(k_f - | \vec{p}+M_N\vec{v}_\sLambda |)}
                {E - m_K - M_N - p^2/(2 \mu_{Kp})}\ ,
\eeq
while that for the $\pi\Sigma$ channel remains unaffected.
Note that we also include the dependence of the \lam\ momentum.  
In Ref.\cite{KochLam} only the \lam\ at rest has been considered. If, as
suggested in \cite{KochLam}, 
the Pauli blocking is responsible for the mass shift of the \lam,
one should expect that for a given density this mass shift should
become small once the momentum of the \lam\ is large compared to the
Fermi-momentum. In this case the momentum distribution of the \lam\
-wave function has very little overlap with other nucleons in the matter and
the effect of the Pauli-blocking becomes small.   

\figampl

This is demonstrated in Fig.~\ref{fig:ampl}. 
There, we show the energy dependence of the calculated scattering amplitude
under the three typical conditions, 
$(\rho,P_\sLambda)=(0,0), (\rho_0/2,0),$ and $(\rho_0/2,0.5 \, {\rm GeV/c})$.
At finite density with the \lam\ at rest, the mean resonance mass 
(around the peak position of the imaginary part) shifts upward
compared with that in free space.
However, once the momentum of \lam\ is large compared with the 
nuclear Fermi momentum,
the effect of the Pauli blocking decreases and the mass shift
is hardly seen.
Therefore, a rather unique signature of the bound state picture of the \lam\ is
a density dependent upwards mass shift which should become smaller as the
momentum of the \lam\ increases with respect to the matter restframe. We are
presently investigating to which
extent this density {\em and} momentum dependent mass shift can be extracted
experimentally by choosing appropriate  kinematic conditions in
$(\gamma, K^+)$, $(\pi, K^+)$ and possibly $(K^-,\pi)$ reactions \cite{KO97}.

\section{Stopped $K^-$ reaction}

The mass shift of \lam\ described above is expected to appear most clearly
if it is created  in the nuclear medium with a small momentum.
The $K^-$ absorption at rest is one possibility,
and another possibility is to use $(K^-,\pi)$ reaction at the magic momentum
of \lam\, at which the \lam\ momentum becomes zero when pions are detected
at forward angles.

In this paper, we focus our attention
to the $K^-$ absorption reaction at rest.
For this reaction, the experimental charged pion spectra are available
for several nuclear targets.
The \lam\ mass shift moves the $I=0$ amplitude  upwards in energy but 
does not affect the  $I=1$ amplitude. It, therefore, modifies 
the relative phase  and strength of $I=0$ and $I=1$ amplitudes
leading to different branching ratios for the reactions 
$K^-p \to \pi^0\Lambda, \pi^-\Sigma^+, \pi^0\Sigma^0, \pi^+\Sigma-$,
as compared to the free ones.

\subsection{Branching ratio}

In Fig.~\ref{fig:brd}, 
we show the model dependence of the calculated branching ratio.
In this figure, we compare the results of the following models or 
parameterizations:
M: Martin's parameterization in free space (without Fermi smearing).
MF: Martin's parameterization in nuclei, with Fermi smearing and binding 
energy correction.
Binding energies are chosen as $B=12.5$ MeV (MFa)~\cite{NOH95}
and $B=17.34$ MeV (MFb).
The latter value corresponds to the neutron separation energy of \nuc{12}{C}.
KF: Koch's amplitude with Fermi smearing, and averaged over density
according to the absorption probability.
The binding energy correction is ignored in the case of KF,
but the calculated pion spectra are not very sensitive
to this binding energy correction, once the mass shift of \lam\ is included.
The density dependence of the branching ratio is also shown in the case of
Koch's amplitude.
The momentum distribution used here has the Gaussian form,
and assumed to be independent on the position.
The same momentum distribution is adopted in all the above calculations.

\figbrd

It is clear that, in the case of KF, the process of $K^-p \to \Sigma^+\pi^-$
is strongly enhanced if  the density is high, 
while the quasi-free $\pi^+$ production ($K^-p \to \Sigma^-\pi^+$) is
suppressed.
As a result, in the above two channels the density averaged branching 
ratios in KF largely differ from those in MF. 
%
%

At a glance, the above difference looks too large, 
since the $K^-$ absorption is considered to occur only at the
nuclear surface region, and therefore, the branching ratio in the low-density
region governs the total (position averaged) ratio.
However, the two models also lead to quite different $K^-$ optical potentials,
and consequently to different absorption point distributions.
In order to illustrate this point, we show the $K^-$ optical potential
and the absorption point distribution in the next two figures.
In Fig.~\ref{fig:potc}, we show  the $K^-$ optical
potential for the two models, MFb and KF, which have been obtained  in
the lowest order impulse approximation 
\beqar
\nonumber
U(r) &=& -{2\pi\hbar^2 \over \mu}\, \left(1+{\mu\over m_{N}}\right)\, \\
\nonumber
	&& \times		\VEV{K^-N|T(\rho(r))|K^-N}_{pn}\, \rho(r)\ \\
\label{eq:opt}
	&& \phantom{MM}		+U_{Coulomb}(r)\ .
\eeqar

As has been already pointed out by Koch~\cite{KochLam}, 
the optical potential of KF has a strong density dependence in accordance with
the analysis of kaonic atoms \cite{FGB93}.
For example, while the $K^-N$ real potential is repulsive at 
low densities, it turns attractive at a finite nucleon density 
($\rho \geq 0.25 \rho_0$).
As a result, the $K^-$ wave function is pulled inside,
towards higher densities.
This is contrary to the straightforward impulse approximation based on the free
T-matrix, which is repulsive in Martin's parameterization. Consequently, 
in this case the $K^-$ wave function is pushed away from the nucleus and 
lower densities are being probed. 

\figpotc

This difference in the radial distribution for the
$K^-$-wave function is shown in   Fig.~\ref{fig:distc}
where the calculated absorption point distribution of
$K^-$ on \C12 is shown.
The absorption probability is assumed to be proportional to 
the imaginary part of the optical potential and the wave function squared,
\beq
        {dP_{abs} \over d\bold{r}}
                \propto - {\rm Im}(U(r))\, |\psi(r)|^2\ ,
\eeq
where $\psi(r)$ is the solution of the Schr\"odinger equation
under the potential $U$.
(We have assumed that $K^-$ is absorbed from a $p$-orbit~\cite{Taka}.)

\figdistc

The calculated average branching ratio can be compared with the one
extracted from experimental data~\cite{Wilquet77},
although there are some assumptions in their estimate, 
such as the conversion probability of $\Sigma$.
Obviously the model KF gives  values closer to the
experimental estimates than models MFa and MFb
in the $I=0$ ($\Sigma\pi$) channels.
Especially, the model KF explains 
the inversion of the branching ratios of
$K^-p \to \Sigma^+\pi^-$ and $\Sigma^+\pi^-$ from free values.

\TABLEA

\subsection{$\pi^-$ spectrum}

In order to calculate the resulting $\pi^-$ spectra 
we have carried out a simple Monte-Carlo simulation
of $K^-$ absorption reaction at rest on \C12. 
This simulation includes
(a) absorption point distribution shown in Fig.~\ref{fig:distc}, 
(b) branching ratio shown in Fig.~\ref{fig:brd},
(c) multiple scattering of the produced pion and hyperon,
(d) $\Sigma$ conversion to $\Lambda$,
and
(e) real part of the optical (mean field) potential for $\Lambda$ and $\Sigma$.
We have assumed that $\pi N$ scattering occurs through the $\Delta$ formation,
and the pion true absorption is described by $\Delta N \to NN$.
The scattering and $\Sigma$ conversion cross sections 
are calculated with Nijmegen model D potential~\cite{NijmegenDYN}.
%
%
The hyperon-nucleus potential is assumed to have the form 
\beqar
        U_Y &=& \alpha_Y {U_N(\rho) \over |U_N(\rho_0)|},\quad
        U_N = \alpha\left({\rho\over\rho_0} \right)
                        +\beta\left({\rho\over\rho_0} \right)^\gamma \ , \\
	\alpha	&=&	-124 \hbox{ MeV} \ ,
	\beta	=	70.5 \hbox{ MeV} \ ,
	\gamma	=	2\ .
\eeqar
The factor $\alpha_\Lambda$= 30 MeV 
is chosen to reproduce the depth of --30 MeV for $\Lambda$ 
in nuclear matter.
For $\Sigma$, we chose $\alpha_\Sigma=$10 MeV.
The collisions between the produced pion and hyperon are ignored.

\figstkcm

In Fig.~\ref{fig:stkcm}, we show the momentum spectrum of $\pi^-$
in \C12(stopped $K^-,\pi^-$) reaction.
The individual contributions are shown in Fig.~\ref{fig:stkcmcmp}.
The highest momentum component comes from the quasi-free $\Lambda$
production,
the component at around 185 MeV/c comes from the weak decay of
$\Sigma^-$ particles,
the largest peak at around 160 MeV/c is due to quasi-free $\Sigma$ production
($K^-p \to \pi^-\Sigma^+, \, K^-n \to \pi^-\Sigma^0$),
and the lowest momentum component comes from the weak $\Lambda$ decay.
The last component ($\Lambda$ decay) includes
the $\Sigma$ conversion and the $\Sigma^0$ electromagnetic decay component
($\Sigma^0 \to \gamma\Lambda$).
All of these yields well agree with the estimate
by Tamura et al.\cite{Tamura89}.

\figstkcmcmp

From Fig.~\ref{fig:stkcm}, 
we see the suppression of the $\Sigma^-$ weak component and the enhancement
of the quasi-free $\Sigma$ production component in KF
compared with the results of MFb.
This is mainly due to the branching ratio change in the model KF.
In addition, since the absorption is assumed to occur only at the surface 
region in the model MFb, $\Sigma$ conversion is estimated to be smaller.
As a result, the model KF agrees better with the experimental
data by Tamura et al.\cite{Tamura89}.

\subsection{$\pi^+$ spectrum}

Since the largest difference in the branching ratio can be seen
in the double charge exchange reaction $K^-p \to \pi^+\Sigma^-$ 
(see Fig.~\ref{fig:brd}), 
a comparison with the $\pi^+$ spectrum should be more sensitive
to the proposed in-medium effect.
Although the quasi-free $\Sigma$ production part of this double
charge exchange reaction is well analyzed by using
direct reaction theories,
the weak decay component of $\Sigma^+$ has been neglected as 
a {\em background}.
Actually, it contains the weak decay of the $\Sigma^+$ particles
with various velocities;
some of them stop in the emulsion (the peak at around 185 MeV/c)
and the rest decay during the slowing down process.
Thus the interpretation of the spectrum is very complicated, and
the extraction of the initial momentum distribution of $\Sigma^+$ particles
is difficult.
%
%
However, we expect that the total yield of $\Sigma^+$ decay
is a rather robust observable.
To demonstrate this, we have attempted to fit the
experimental data~\cite{Kubota96},
including the `background' from the weak $\Sigma^+$ decay,
by taking into account the following sources.
We have assumed that the weak decay spectra of stopped and moving $\Sigma^+$
centered at 185 MeV/c have a Lorenzian and two-range Gaussian form,
respectively.
The $\pi^+$ free production on hydrogen at 173 MeV/c is assumed to be
a Lorenzian,
and a Breit-Wigner type spectrum with threshold correction is used to fit
the quasi-free part on \C12.
The charge exchange reactions such as $K^-N \to \pi \Lambda$ followed by
$\pi N \to \pi^+ N'$ are ignored in the fitting procedure,
since their contributions are negligible.
The resulting fit is shown in Fig.~\ref{fig:expfit}.



By using this fit, we can reconstruct the $\pi^+$ spectrum from
the $K^-$ absorption on \C12, where the subsequent atomic processes
such as slowing down of $\Sigma^+$ are removed.
For this reconstruction,
the following free branching ratios are used;
$K^-p \to \pi^-\Sigma^+ (4.37 \%), \pi^+\Sigma^- (28.66 \%)$
and $\Sigma^+ \to \pi^+ n (48.3 \%)$.
In addition, the $\Sigma^+$ weak decay spectrum without the subsequent 
atomic processes is assumed to have the same shape as
the fitted moving $\Sigma^+$ spectrum.
In Fig.~\ref{fig:stkcp}, 
we show the calculated $\pi^+$ spectra with this reconstructed data.
We have adjusted the scale of this reconstructed data
by requiring that its weak decay tail
agrees with the calculated results approximately,
then the difference at the quasi-free part becomes significant:
The yield of the quasi-free component in the model MFb is 
larger by a about a factor of two than that in the model KF.


This difference mainly comes from the branching ratio difference
in the $K^-p \to \pi+\Sigma^-$ channel.
The absolute value of the $\Sigma^+$ particles which decay outside 
of the nucleus is similar in the two models, 
because the reaction occurs at the surface in the model MFb and 
$\Sigma^+$ particle can easily escape from nuclei.
However, the relative yield of $\Sigma^+$ particles as compared 
to the quasifree peak, which essentially is
the ratio of $\Sigma^+ / \Sigma^-$, differs by more than a factor of two. 


In order to make a clear comparison, 
we have extracted an approximate value for the branching ratio,
by taking the ratio of the contribution to the quasi-free bump
to the sum of moving $\Sigma^+$ and $\Sigma^+$ at rest.
This can then be directly compared with the results
from the different models (see Table~\ref{table:multi}).
Clearly, the model based on the $K^-p$-bound state 
for the \lam\ compares best
with our estimate of the branching ratio from experiment.
It is still 50 \% off, though,
but as compared to the free space value 
or that including Fermi smearing, it constitutes a significant improvement.

The remaining difference may very well be due to 
the interaction between $\Sigma$ and the residual nuclei, $^{11}$B$^*$,
which can have a structure $\alpha+\alpha+t$.
It is well known that the $\Sigma N$ interaction has a strong spin-isospin
dependence, and there is a large uncertainty in its strength.
In addition, when resonance states or unstable bound states such as 
$^4_\Sigma$He are created in the subsystem,
the $\Sigma$ conversion probability as well as the initial branching
ratio may be modified from the present estimate.
All of these effects are ignored in this work,
and should be included in future works,
once they are reasonably well controlled.

\figexpfit

\figstkcp

\TABLEB

\section{Summary and outlook}
\label{sec:summary}

In this paper, we consider two extreme pictures of the \lam\ and their
consequences for stopped $K^-$ experiments.
The first scenario assumes that \lam\ is a three-quark state,
rather than a bound state of $\bar{K}N$.
In this case it will not be subject to Pauli blocking in the nuclear medium,
and the matrix elements such as $K^-N \to \pi\Sigma$ are density independent.
In the other scenario, 
we have assumed that \lam\ is a bound state of $\bar{K}N$ and, thus, is
subject to Pauli-blocking in the medium. While the full, Pauli corrected
$T$-matrix has been used in the calculations, the dominant effect is a density
and momentum dependent mass shift of the \lam.
This mass shift modifies the relative phase and relative strength between 
the matrix element of $I=0$ and $I=1$ channels,
and leads to a change of the  branching ratio.
It also affects the potential between $K^-$ and the target nucleus,
and as a result, $K^-$ absorption can occur deeply inside the nucleus
in the latter scenario.

We have simulated the $K^-$ absorption reaction at rest on \C12 nucleus
by combining these results.
The largest difference between the results of two scenarios
can be found in the ratio $\Sigma^+\pi^-/\Sigma^-\pi^+$.
In the bound state picture,
the first channel is enhanced at finite densities
by the interference of $I=0$ and $I=1$ amplitudes.
This difference appears most clearly in the $\pi^+$ spectrum.
The experimental data~\cite{Kubota96} suggest
that the escaping $\Sigma^+$ particle is more abundant than
$\pi^+$ particles produced in the quasi-free process, 
$K^-p \to \Sigma^-\pi^+$.
The comparison of the calculated results of $\pi^-$ and $\pi^+$ spectra 
with the experimental data~\cite{Tamura89,Kubota96}
suggests that
the latter scenario --- the \lam\ is a bound state of $\bar{K}N$ ---
gives a better description of this reaction,
due to the enhancement of $K^-p \to \Sigma^+\pi^-$ branching ratio.
Therefore, we conclude that the dominant component of the \lam\ wave function 
is the $\bar{K}N$ bound state.

The above conclusion is, however, only  an in-direct support of the latter 
scenario, since the change in the branching ratio is mostly due to the mass
shift of the \lam.
Such a  mass shift could for instance also be due to simple
mean field effects. Also, it would be of course preferable if one could 
observe the mass shift directly.
The in-medium effect which is more unique to the bound state picture is the
momentum dependence of the mass shift of the  \lam;
if the \lam\ momentum is larger than the nuclear Fermi momentum,
the effects of the Pauli blocking is reduced, and the mass shift disappears.
Therefore, the best confirmation for the bound state picture of the \lam\
will be the observation of both the momentum and density  dependence 
of its mass shift,
by comparing two kind of reactions 
---     the first involves "fast" \lam\ particles at finite nuclear densities,
        and in the second type of reactions,
	\lam\ particles are produced at very small momenta.
As for the second type of reactions, it is the most ideal to use the magic
momentum of \lam\ in $(K^-,\pi^-)$ reaction.
This reaction can be used for the first, too,
when $\pi^-$ particle is detected at relatively large angles.
The work on this line is in progress \cite{KO97}.

\bigskip

The authors would like to thank
Dr. H. Tamura, Dr. H. Outa, Professor T. Harada, Professor K. Yazaki,
Mr. T. Koike, Dr. J. Schaffner and Dr. M. Arima for fruitful discussions.
This work was supported in part by
the Japan Society for Promotion of Science, 
the Grant-in-Aid for Scientific Research (Nos.\ 07640365 and 09640329)
and the Overseas Research Fellowship
from the Ministry of Education, Science and Culture, Japan,
and by the Director, Office of Energy Research,
Office of High Energy and Nuclear Physics,
Nuclear Physics Division of the U.S. Department of Energy
under Contract No.\ DE-AC03-76SF00098.
Two of the authors (A.O. and Y.N.) also thank the members of nuclear
theory group for their hospitality during the stay at LBNL.

%
\ifx\undefined\ifEPRINT \newpage \fi

\ifx\undefined\ifBIBTEX

\else
       \bibliography{hyper}
\fi

        \ifx\undefined\ifEPRINT
\Figampl
\Figbrd
\Figpotc
\Figdistc
\Figstkcm
\Figstkcmcmp
\Figexpfit
\Figstkcp
\TableA
\TableB
        \fi

\WIDETEXT
\end{document}